\documentstyle[preprint,aps,epsf,floats]{revtex} 

\def\lchi{\Lambda_\chi}
\def\ket#1{\left| #1 \right\rangle}
\def\bra#1{\left\langle #1 \right|}
\def\msbar{$\overline{\rm MS}$}

\begin{document}

\preprint{\tighten \vbox{\hbox{UCSD/PTH 99-22}}}

\title{On the structure of large $N_c$ cancellations in baryon chiral
perturbation theory}

\author{Ruben Flores-Mendieta, Christoph P.~Hofmann, Elizabeth Jenkins,\\
and Aneesh V.~Manohar}

\address{Department of Physics, University of California at San Diego, La
Jolla, CA 92093}

\maketitle

{\tighten
\begin{abstract}%
We show how to compute loop graphs in heavy baryon chiral perturbation theory
including the full functional dependence on the ratio of the Delta--nucleon
mass difference to the pion mass, while at the same time automatically
incorporating the $1/N_c$ cancellations that follow from the large-$N_c$
spin-flavor symmetry of baryons in QCD.  The one-loop renormalization of the 
baryon axial vector current is studied to demonstrate the procedure.  A new
cancellation  is identified in the  one-loop contribution to the baryon axial
vector current.  We show that loop corrections to the axial vector currents are
exceptionally sensitive  to deviations of the ratios of baryon-pion axial
couplings from $SU(6)$ values.
\end{abstract}
}

\newpage


\section{Introduction} 

Baryon chiral perturbation theory can be used to systematically compute the
properties of baryons as a function of the light quark masses $m_q$. A physical
quantity's non-analytic dependence on $m_q$ is calculable from pion-loop
graphs; its analytic dependence has contributions both from pion-loop graphs
and from low-energy constants that are present in the chiral Lagrangian.

It is convenient to formulate baryon chiral perturbation theory in terms of
velocity-dependent baryon fields, so that the expansion of the baryon chiral
Lagrangian in powers of $m_q$ and $1/m_B$ (where $m_B$ is the baryon mass) is
manifest~\cite{Jenkins:1991ne,Jenkins:1991jv}.   This formulation is called
heavy baryon chiral perturbation theory.   The earliest application of heavy
baryon chiral perturbation theory was to baryon axial vector
currents~\cite{Jenkins:1991jv}.   Two important results were obtained from this
analysis.  First, the baryon axial coupling ratios were found to be close to
their $SU(6)$ values with an $F/D$ ratio close to $2/3$, the value predicted by
the non-relativistic quark model.  Second, there were large cancellations in the
one-loop corrections to the baryon axial vector currents between loop graphs
with intermediate spin-$1/2$ octet and spin-$3/2$ decuplet baryon states. It
was later proven using the $1/N_c$ expansion that the baryon axial couplings
ratios should have $SU(6)$ values with $F/D=2/3$, up to corrections of order
$1/N_c^2$ for pions~\cite{Dashen:1993as,Dashen:1994jt}.\footnote{The ratio
$F/D$ is equal  to $2/3$ in the quark representation, and to $5/8$ in the
Skyrme representation.   The quark and Skyrme representations are equivalent
in large $N_c$~\cite{Manohar:1984md} up
to corrections of relative order  $1/N_c^2$ for pions~\cite{Dashen:1993as}.  We
will use the quark representation in this paper.} In addition, it was shown
that axial vector current loop graphs with octet and decuplet intermediate
states cancel to various orders in $N_c$.  For nucleon and Delta intermediate
states, there is a cancellation of the one-loop graphs to two orders in $N_c$;
each individual one-loop diagram is of order $N_c$, but the sum of all
one-loop  diagrams is of order $1/N_c$~\cite{Dashen:1993as}. Similar
large-$N_c$ cancellations also occur for other baryonic 
quantities~\cite{Jenkins:1991ne,Butler}.

We would like to find a calculational scheme that simultaneously exhibits both
the $m_q$ and $1/N_c$ expansions. In the chiral limit $m_q \to 0$, pions
become massless Goldstone boson states. There is an expansion about the chiral
limit in powers of $m_q/\lchi$, or equivalently, in powers of
$m_\pi^2/\lchi^2$, where $\lchi\sim 1~{\rm GeV}$ is the scale of chiral
symmetry breaking and $m_\pi$ is the pion
mass.\footnote{For $SU(3)_L\times SU(3)_R$ chiral symmetry, there are loop
corrections involving the pions, kaons and $\eta$ which depend on the pion,
kaon and $\eta$ masses, respectively. Chiral perturbation theory depends on the
expansion parameters $m_\pi^2/\lchi^2$,  $m_K^2/\lchi^2$ and
$m_\eta^2/\lchi^2$.  Large-$N_c$ chiral perturbation theory also depends on the $\eta^\prime$
mass.} In the large-$N_c$ limit, the nucleon and Delta become
degenerate, $\Delta\equiv M_\Delta -M_N \propto 1/N_c \to 0$, and form a single
irreducible representation of the contracted spin-flavor symmetry of baryons in
large-$N_c$ QCD~\cite{Dashen:1993as,Dashen:1994jt}. There is an
expansion in powers of $1/N_c$ about the large-$N_c$ limit.  We will consider a
combined expansion in $m_q/\lchi$  and $1/N_c$ about the double limit $m_q
\rightarrow 0$ and $N_c \rightarrow \infty$.

Loop graphs in heavy baryon chiral perturbation theory have a calculable
dependence  on the ratio $m_\pi/\Delta$. In general, this dependence is
described by a function $F(m_\pi,\Delta)$. In the chiral limit $m_q \to 0$ with
$\Delta$ held fixed,  the function can be expanded in powers of $m_\pi/\Delta$,
\begin{equation}\label{eq1}
F\left(m_\pi ,\Delta\right) = F_0 + \left({m_\pi \over \Delta}\right) F_1
+ \left({m_\pi \over \Delta}\right)^2 F_2 + \ldots,
\end{equation}
whereas in the $1/N_c \to 0$ limit with $m_\pi$ held fixed,  the function can
be expanded in powers of $\Delta/m_\pi$,
\begin{equation}\label{eq2}
F\left(m_\pi, \Delta\right) = \bar F_0 + 
\left({\Delta \over m_\pi }\right) \bar F_1
+\left({\Delta \over m_\pi }\right)^2 \bar F_2 + \ldots\ .
\end{equation}
The difference between the two expansions in Eqs.~(\ref{eq1}) and~(\ref{eq2})
is commonly referred to as the non-commutativity of the chiral and large-$N_c$
limits~\cite{cohen}. 

It is important to remember, however, that the conditions for heavy baryon
chiral perturbation theory (including Delta states) to be
valid are that $m_\pi \ll \lchi$ {\bf and} $\Delta \ll \lchi$.  The ratio
$m_\pi/\Delta$ is not constrained and can take any value.
The entire dependence of a physical quantity on
$m_\pi/\Delta$ is {\bf calculable} in heavy baryon chiral perturbation
theory~\cite{Jenkins:1991ne}, so the ratio $m_\pi/\Delta$ need not be small or large
for calculations. 
In the real world,  $m_\pi/\Delta \sim 0.5$, so
it is useful to have a calculational scheme that retains the full functional
dependence of $F(m_\pi,\Delta)$ on the ratio $m_\pi/\Delta$. A straightforward
approach is to simply calculate the full dependence on $m_\pi/\Delta$ of the
loop graphs, and evaluate the loop correction at the physical value
$m_\pi/\Delta \sim 0.5$~\cite{Jenkins:1991ne,cohen2}. Another common procedure 
advocated in
the literature is to not include intermediate Delta particles explicitly in
loops, but to incorporate their effects into the low-energy constants of the
effective Lagrangian~\cite{meissner}. The disadvantage of this second approach
is that one finds large numerical cancellations between loop diagrams with
intermediate nucleon states and low-energy constants containing  the effects of
Delta states. These cancellations are guaranteed to occur as a consequence of
the contracted spin-flavor symmetry which is present in the $N_c \to \infty$
limit.
The large-$N_c$ spin-flavor symmetry responsible for the
cancellations is hidden in this approach because including only the spin-$1/2$ baryons 
in the chiral Lagrangian breaks the large-$N_c$ spin-flavor symmetry explicitly, 
since the spin-$1/2$
and spin-$3/2$ baryons together form an irreducible representation of
spin-flavor symmetry. 
Because the sum of the loop
contributions with intermediate octet and decuplet states respects spin-flavor
symmetry and is much smaller (by powers of $1/N_c$) than each individual loop
contribution separately, it is important to keep the large-$N_c$ spin-flavor
symmetry of the baryon chiral Lagrangian and the large-$N_c$ cancellations
manifest.

In this paper, we will show how one can combine heavy baryon chiral perturbation
theory with the $1/N_c$ expansion so that the full-dependence on $m_\pi/\Delta$
is retained and the $1/N_c$ cancellations are explicit.  This method has the
advantage that the loop correction to the baryon axial isovector current, which
is order $1/N_c$, is automatically obtained to be of this order, instead of as
the sum of two contributions (loop correction and counterterm) of order $N_c$
which cancel to two powers in $1/N_c$.  Note that at higher orders  the
cancellations become more severe, and it is even more important to keep the
$1/N_c$ cancellations manifest.  For example, at two loops, each loop diagram
is naively of order  $N_c^2$, whereas the sum of all two-loop diagrams is order
$1/N_c^2$.  Not including the  $1/N_c$ cancellations in a systematic way gives
a misleading picture of the baryon chiral expansion---one finds higher order
corrections that grow with $N_c$, which is incorrect.  Including the $1/N_c$
cancellations restores the $1/N_c$ power counting so that the loop corrections
are suppressed by the factor $1/N_c^L$, where $L$ is the number of loops.

The organization of this paper is as follows. In Sec.~\ref{sec:overview},  we
begin with a brief overview of the $1/N_c$ cancellations occurring in the
one-loop correction to the baryon axial vector currents. In
Sec.~\ref{sec:formula}, we derive the formula for the one-loop correction to
the baryon axial vector currents for arbitrary $\Delta/m_\pi$, in a form that
is convenient for later use. The structure of large-$N_c$ cancellations for
$\Delta/m_\pi=0$ and $\Delta/m_\pi \not=0$ are discussed in
Secs.~\ref{sec:cancel} and \ref{sec:nonzero}, respectively.  The general power
counting for large-$N_c$ cancellations is derived to all orders in the baryon
hyperfine mass splitting $\Delta$, and it is determined that the dominant
large-$N_c$ cancellations are present only  in terms that are of low and finite
order in $\Delta$.  A procedure for subtracting and isolating  these
large-$N_c$ cancellations is given.  Other contributions to axial vector
current  renormalization are briefly presented in Sec.~VI.  Our conclusions are
summarized in Sec.~VII. 

\section{Overview}\label{sec:overview}

A brief review of heavy baryon chiral perturbation theory and the $1/N_c$
baryon chiral Lagrangian can be found in Ref.~\cite{fhj}, so only a few salient
facts will be repeated here.  The pion-baryon vertex is proportional to
$g_A/f$, where $f$ is the decay constant of the $\pi$ meson.  In the
large-$N_c$ limit, $g_A \propto N_c$ and $f \propto \sqrt{N_c}$, so that the
pion-baryon vertex is of order $\sqrt{N_c}$ and grows with $N_c$.  The baryon
propagator  is $i/(k\cdot v)$ and is $N_c$-independent, as is the pion
propagator. In the \msbar\ scheme, all loop integrals are given by the pole
structure of the propagators, so loop integrals do not  depend on $N_c$.

\begin{figure}
\epsfxsize=15cm
\hfil\epsfbox{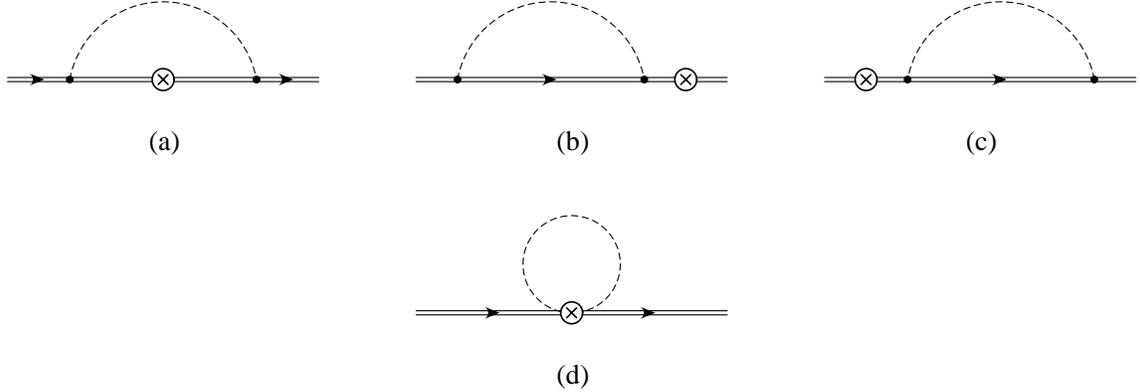}\hfill
\caption{One-loop corrections to the baryon axial vector current.
\label{fig:axial}}
\end{figure}

The tree-level matrix element of the baryon axial vector current is of order
$N_c$, since $g_A$ is of order $N_c$. The one-loop diagrams that renormalize
the baryon axial vector current are shown in Fig.~\ref{fig:axial}.
Each of the one-loop corrections in Fig. 1(a,b,c) involves two pion-baryon
vertices, and is order $N_c$ times the tree-level graph. 

The matrix elements of the space components of the baryon axial vector current
between initial and final baryon states $B$ and $B^\prime$ will be denoted by
\begin{equation}
\bra{B^\prime} \bar \psi \gamma^i \gamma_5 T^a \psi \ket{B} 
= \left[A^{ia} \right]_{B^\prime B},
\end{equation}
where $B$ and $B^\prime$ are baryons in the lowest-lying irreducible
representation of  contracted-$SU(6)$ spin-flavor symmetry, i.e. the spin-$1/2$
octet and spin-$3/2$ decuplet baryons. The Feynman diagram amplitude for $B \to
B^\prime + \pi (k)$ is $\left[A^{ia} \right]_{B^\prime B} {\mathbf k}^i/f$,
where $\bf k$ is the three-momentum of the emitted pion. The time-component of
the axial current has zero matrix element between static baryons, and is
represented in the heavy baryon formulation by a higher dimension operator in
the effective Lagrangian.  The matrix elements 
$\left[A^{ia} \right]_{B^\prime B}$ of the spatial components of the axial vector
current can be written in
terms of the octet and decuplet pion coupling constants $F$, $D$, $\cal C$, and
$\cal H$~\cite{Jenkins:1991jv}, each of which is of order $N_c$. 

The one-loop correction to the baryon axial vector current, in the limit that the
Delta-nucleon mass difference is neglected, is proportional to the double
commutator
\begin{equation}\label{30}
\delta A^{ia} \propto {1\over f^2} 
\left[A^{jb},\left[A^{jb},A^{ia}\right]\right],
\end{equation}
where the sum over intermediate baryon states is given by matrix multiplication
of the $A^{ia}$ matrices. Naively, the double commutator is of order $N_c^3$,
and $f \propto \sqrt{N_c}$ so that $\delta A^{ia}$ is of order $N_c^2$. One of
the results of the $1/N_c$ analysis for baryons is that the double commutator
is of order $N_c$, rather than  $N_c^3$~\cite{Dashen:1993as}. Each individual
term in the sum Eq.~(\ref{30}) is of order $N_c^3$, but there is a cancellation
in the sum over intermediate baryons, which is guaranteed by the spin-flavor
symmetry of large-$N_c$ QCD~\cite{Dashen:1993as,Dashen:1994jt}. The
cancellation only occurs when the ratios of $F$, $D$, $\cal C$, and $\cal H$
are close to their $SU(6)$ values.\footnote{An important point to note is that
large-$N_c$ QCD predicts only the ratios of $F/ D$, ${\cal C}/D$, and ${\cal
H}/D$, the overall normalization of the coupling constants is not fixed by the
symmetry.  The large-$N_c$ cancellations depend on the coupling ratios being
close to their $SU(6)$ values, and do not depend on the overall  normalization
of the couplings.} The large-$N_c$ cancellation implies that the one-loop
correction to the axial current is $1/N_c$ times the tree-level value, instead
of $N_c$ times the tree-level value. Similarly, the two-loop correction is
$1/N_c^2$ times the tree-level value, instead of $N_c^2$ times the tree-level
value (see Ref.~\cite{Keaton:1996ak} for an explicit calculation in the degeneracy
limit). The one-loop
large-$N_c$ cancellations will be discussed more fully in Secs.~IV and~V.  The
formalism for making large-$N_c$ cancellations manifest  is provided in Sec.~V.

The large-$N_c$ cancellation in the one-loop correction to the baryon axial
vector current can be seen numerically from explicit computation in heavy
baryon chiral perturbation theory. The baryon axial vector current matrix
element at one-loop has the form
\begin{equation}
A = \alpha + \left(\bar \beta - \bar \lambda \alpha\right) 
{m^2 \over 16 \pi^2 f^2} \ln {m^2 \over \mu^2} + \ldots
\end{equation}
where $\alpha$ is the tree-level contribution, $\bar\beta$ is the vertex
correction, $\bar\lambda$ is the wavefunction renormalization, $m$ is the
$\pi$,  $K$ or $\eta$ mass, and the Delta-nucleon mass difference has been
neglected for simplicity so there is  only a chiral-logarithmic contribution.
[The full one-loop correction will be discussed in the next section.] For the
case of  $\bra p \bar u \gamma^\mu \gamma_5 d \ket n$, the coefficients are
\begin{eqnarray}
\alpha &=& D + F \, ,\nonumber \\
{\bar \lambda}_\pi &=& \frac94 {(F + D)}^2 + 2 {\cal C}^2 \, ,\nonumber \\
{\bar \lambda}_K &=& \frac12 (9 F^2 - 6 F D + 5 D^2 + {\cal C}^2)\, ,\nonumber \\
{\bar \lambda}_\eta &=& \frac14 {(3 F - D)}^2 \, ,\nonumber \\
{\bar \lambda}_{\eta^\prime} &=&  2D^2\, ,\nonumber \\
{\bar \beta}_{\pi} & = & \frac14 {(F + D)}^3 + \frac{16}{9}
(F + D) {\cal C}^2 - \frac{50}{81} {\cal H} {\cal C}^2 - F - D \, ,
\nonumber \\
{\bar \beta}_{K} & = & \frac13 (-3 F^3 + 3 F^2 D - F D^2 +
D^3) + \frac29 (F + 3 D) {\cal C}^2 - \frac{10}{81} {\cal H} {\cal C}^2
- \frac12 (F + D) \, , \nonumber \\
{\bar \beta}_{\eta} & = & - \frac{1}{12} (F + D){(3 F - D)}^2
\, , \nonumber \\
{\bar \beta}_{\eta^\prime} & = & - \frac{1}{12} (F + D){(3 F - D)}^2 \, . 
\end{eqnarray}
The coefficients for the other matrix elements can be found in the  literature
\cite{Jenkins:1991jv}. The subscripts $\pi$, $K$ and $\eta$ denote the
contributions from $\pi$, $K$ and $\eta$ loops. To illustrate the cancellation,
we have plotted the one-loop coefficients $\left(\bar \beta - \bar \lambda
\alpha\right)$ 
\begin{figure}
\epsfxsize=10cm
\hfil\epsfbox{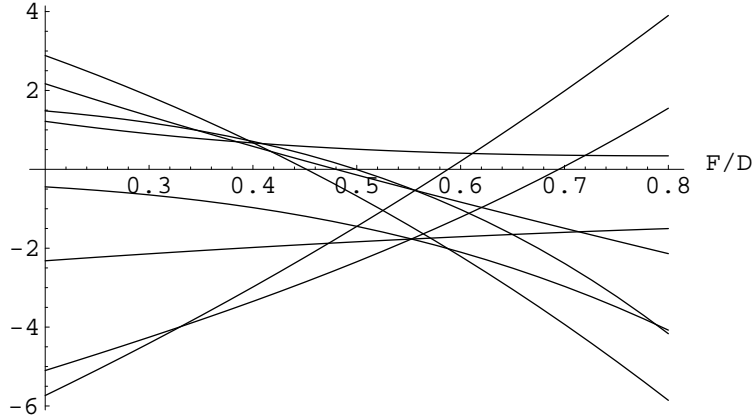}\hfill
\caption{One-loop pion correction to the baryon axial vector current  $\bra p
\bar u \gamma^\mu \gamma_5 d \ket n$. The curves are $\bar \beta-\bar \lambda
\alpha$ for (from top to bottom along the left hand edge of the graph) $N \to N
\pi$, $\Xi\to \Lambda \bar K$, $\Sigma \to \Sigma \pi$, $\Xi \to \Xi \pi$, $\Xi
\to \Sigma \bar K$, $\Sigma \to \Lambda \pi$, $\Sigma \to N \bar K$, $\Lambda
\to N \bar K$. 
\label{fig:picancel}}
\end{figure}%
\begin{figure}
\epsfxsize=10cm
\hfil\epsfbox{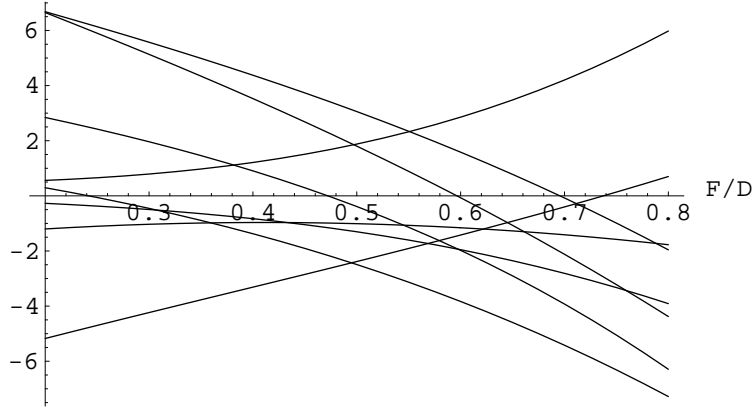}\hfill
\caption{One-loop kaon correction to the baryon axial current  $\bra p \bar u
\gamma^\mu \gamma_5 d \ket n$. The curves are $\bar \beta-\bar \lambda \alpha$
for (from top to bottom along the left hand edge of the graph) $\Xi \to \Xi
\pi$, $\Sigma \to \Sigma \pi$, $\Xi\to \Lambda \bar K$, $\Lambda \to N \bar K$,
$\Xi \to \Sigma \bar K$, $N \to N \pi$, $\Sigma \to \Lambda \pi$, $\Sigma \to N
\bar K$.
\label{fig:Kcancel}}
\end{figure}%
\begin{figure}
\epsfxsize=10cm
\hfil\epsfbox{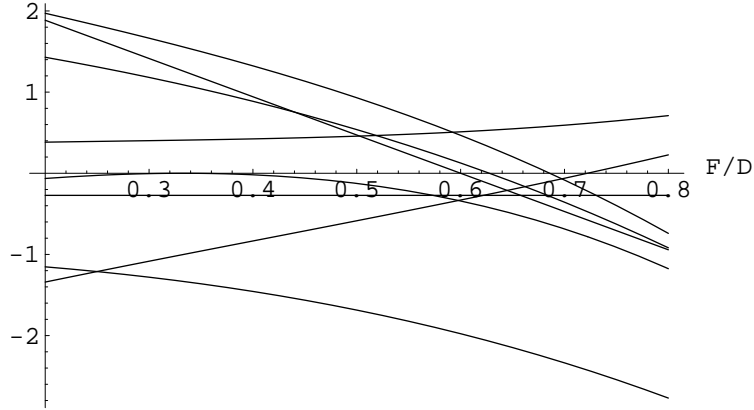}\hfill
\caption{One-loop eta correction to the baryon axial current  $\bra p \bar u
\gamma^\mu \gamma_5 d \ket n$. The curves are $\bar \beta-\bar \lambda \alpha$
for (from top to bottom along the left hand edge of the graph) $\Xi \to \Xi
\pi$, $\Sigma \to \Sigma \pi$, $\Xi\to \Lambda \bar K$, $\Lambda \to N \bar K$,
$N \to N \pi$, $\Sigma \to \Lambda \pi$, $\Xi \to \Sigma \bar K$, $\Sigma \to N
\bar K$.
\label{fig:etacancel}}
\end{figure}%
for the axial currents (or equivalently, the couplings) $N \to N \pi$, $\Sigma
\to \Lambda \pi$, $\Sigma \to \Sigma \pi$, $\Xi \to \Xi \pi$, $\Lambda \to N
\bar K$, $\Sigma \to N  \bar K$,  $\Xi \to \Lambda \bar K$, and $\Xi \to \Sigma
\bar K$ in Figs.~\ref{fig:picancel}, \ref{fig:Kcancel}, and
\ref{fig:etacancel}.  For simplicity, the coefficients are plotted as a
function of $F/D$ only --- the  other coupling ratios have been fixed at their
$SU(6)$ values ${\cal C}/D=-2$  and ${\cal H}/D=-3$. The best fit to the baryon
axial currents has the axial coupling ratios close to their $SU(6)$
values~\cite{Jenkins:1991ne,Jenkins:1991jv}, so this is a reasonable approximation. The
large-$N_c$ analysis indicates that there should be some cancellation in the
loop correction when $F/D$ is close to the $SU(6)$ value of $2/3$. This
suppression is evident separately for the $\pi$, $K$ and $\eta$ loops for all
eight processes. This is the cancellation pointed out phenomenologically in
Ref.~\cite{Jenkins:1991ne,Jenkins:1991jv}, and later proved in
Refs.~\cite{Dashen:1993as,Dashen:1994jt}.  We will study this cancellation
quantitatively in terms of the $1/N_c$ expansion in this work.

\section{One-loop correction to the axial current}\label{sec:formula}

The one-loop diagrams that contribute to the baryon axial vector current are
shown in Fig.~\ref{fig:axial}. Figs.~\ref{fig:axial}(a,b,c) are of order $N_c$
times the tree-level vertex, and Fig.~\ref{fig:axial}(d) is of order $1/N_c$
times the tree-level vertex. The large-$N_c$ cancellations occur between 
Figs.~\ref{fig:axial}(a,b,c), so we will concentrate on these three diagrams in
this section. The contribution from Fig.~\ref{fig:axial}(d) is considered
briefly in Sec. VI.  Both contributions can be found in
Ref.~\cite{Jenkins:1991jv}.\footnote{Fig.~\ref{fig:axial}(d)  is linear in the
pion-baryon coupling constants $F$, $D$ and $\cal C$, whereas
Figs.~\ref{fig:axial}(a,b,c) are cubic, so it is easy to identify the two
pieces in existing calculations.}

All the loop graphs we need can be written in terms of the basic loop integral
\begin{equation}\label{Fdef}
\delta^{ij}\, F(m,\Delta,\mu) = {i \over f^2} \int {{d^4k} \over {(2 \pi)^4}}
{{({{\bf k}^i})(-{{\bf k}^j})} \over 
{(k^2 - m^2)(k\cdot v- \Delta+i\epsilon)} },
\end{equation}
where $\mu$ is the scale parameter of dimensional regularization.  Evaluating
the integral gives
\begin{eqnarray}
{24 \pi^2 f^2} \ F(m, \Delta, \mu)&& \nonumber \\
&&= \Delta \left(\Delta^2 -{3 \over 2} m^2 \right)
\ln{m^2 \over \mu^2}- {8 \over 3} \Delta^3-{7 \over 2}\Delta
m^2\nonumber\\[10pt]
&&\qquad +\cases{
2 \left( m^2 - \Delta^2 \right)^{3/2} \left[ {\pi \over 2} - {\rm tan}^{-1}
\left( {\Delta \over \sqrt{m^2 - \Delta^2}}\right) \right],
& $\left|\Delta\right| \le m$,\cr\noalign{\medskip}
- \left( \Delta^2 - m^2 \right)^{3/2}
\ln \left( { {\Delta - \sqrt{\Delta^2 - m^2}} \over
{\Delta + \sqrt{\Delta^2 - m^2}}} \right)
,& $\left|\Delta\right| > m$ .\cr}\label{F}
\end{eqnarray}

\subsection{Wavefunction Renormalization}

\begin{figure}
\epsfxsize=5cm
\hfil\epsfbox{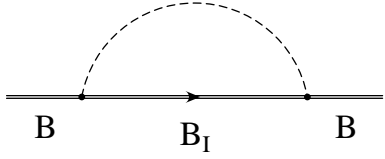}\hfill
\caption{One-loop wavefunction renormalization graph.
\label{fig:wave}}
\end{figure}

The wavefunction renormalization graph for baryon $B$ is shown in
Fig.~\ref{fig:wave}, where one sums over all possible intermediate baryons $B_I$.
The loop graph is equal to
\begin{eqnarray}
i G_{B} & =& \sum_{j,k,b,B_I}{i^2 \over f^2} \left[A^{kb} \right]_{B B_I}
\left[A^{jb} \right]_{B_I B}\int {{d^4k} \over {(2 \pi)^4}}
{{({{\bf k}^k})(-{{\bf k}^j})} \over 
{\left(k^2 - m^2_b \right)\left( \left(k+p \right)\cdot v  - 
\left(M_I-M \right) +i\epsilon \right)} }, \label{2}
\end{eqnarray}
where $b=1,\dots,9$ or $\pi,K,\eta,\eta^\prime$ labels the intermediate
meson.\footnote{The $\eta^\prime$ is a ninth Goldstone boson in the large $N_c$
limit~\cite{etaprime,pich}. Our formulae apply to the $\eta^\prime$ corrections, with
flavor matrix $\lambda^9=\sqrt{2/3}$.  The formalism for including the $\eta^\prime$ 
is described
in detail in Ref.~\cite{fhj}.} The wavefunction renormalization correction for
baryon $B$ is
\begin{equation}\label{3}
Z_{B} = -\left. {\partial G_B \over \partial (p \cdot v)} \right|_{p\cdot v=0}.
\end{equation}
The wavefunction correction to the axial vector current matrix element $\bra{B^\prime} A^{ia} \ket{B}$
depends on
\begin{eqnarray}
Z_{B^\prime B} = {1 \over 2} \left( Z_{B^\prime} + Z_B \right),
\end{eqnarray}
which can be written in terms of the function $F(m,\Delta,\mu)$ defined in
Eq.~(\ref{Fdef}) as
\begin{equation}\label{4}
Z_{B^\prime B} = \sum_{j,b,B_I}\left[A^{jb} \right]_{B^\prime B_I}\left[A^{jb}\right]_{B_I B} 
{\partial F(m_b,\Delta_{B_I B},\mu) \over \partial \Delta_{B_I B}} ,
\end{equation}
where
\begin{equation}\label{5}
\Delta_{B_1 B_2}\equiv M_{B_1}-M_{B_2}.
\end{equation}
The wavefunction renormalization correction $Z_{B^\prime B}$ is diagonal in
flavor and spin.

\subsection{Vertex Correction}

\begin{figure}
\epsfxsize=5cm
\hfil\epsfbox{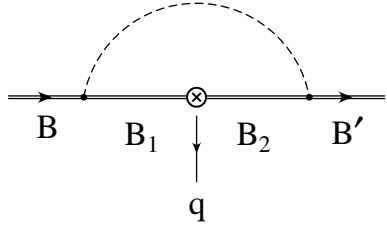}\hfill
\caption{One-loop vertex correction graph.
\label{fig:vertex}}
\end{figure}

The one-loop correction to the matrix element $\bra{B^\prime} A^{ia} \ket{B}$ from the vertex
graph Fig.~\ref{fig:vertex} can be written as
\begin{eqnarray}
&&\left[\delta A^{ia} \right]^{\rm vertex}_{B^\prime B} = 
\sum_{j,k,b,B_1,B_2}-{i \over f^2} \left[A^{kb} \right]_{B^\prime B_2}
\left[A^{ia} \right]_{B_2 B_1}\left[A^{jb} \right]_{B_1 B}\nonumber \\
&&\times \int {{d^4k} \over {(2 \pi)^4}}
{{({{\bf k}^k})(-{{\bf k}^j})} \over 
{\left(k^2 - m^2_b \right)\left( k \cdot v  - 
\left(M_{1}-M \right) +i\epsilon \right)\left( \left(k-q\right) \cdot v  - 
\left(M_{2}-M \right) +i\epsilon \right)} },\label{6}
\end{eqnarray}
where $q$ is the outgoing momentum transfer at the axial vertex. For octet-octet matrix
elements, $q\cdot v =0$, whereas for decuplet-octet transition matrix elements, $q
\cdot v=M-M^\prime$, the average decuplet-octet mass difference. One can rewrite
the denominator of Eq.~(\ref{6}) using the identity
\begin{eqnarray}\label{7}
{1 \over {(k^0- \Delta_1+i\epsilon)(k^0- \Delta_2+i\epsilon)} }
= {1 \over {(\Delta_1 - \Delta_2)}} 
\left[ {1 \over {(k^0- \Delta_1+i\epsilon)}} -
{1 \over {(k^0- \Delta_2+i\epsilon)}}
\right]
\end{eqnarray}
so that
\begin{eqnarray}
\left[\delta A^{ia} \right]^{\rm vertex}_{B^\prime B} & =& 
-\sum_{j,b,B_1,B_2} \left[A^{jb} \right]_{B^\prime B_2}
\left[A^{ia} \right]_{B_2 B_1}\left[A^{jb} \right]_{B_1 B}\nonumber \\
&&\times 
{1\over  \Delta_{B_1 B}-  \Delta_{B_2 B^\prime} }\left[ F\left(m_b,\Delta_{B_1 B},\mu\right)-
F\left(m_b,\Delta_{B_2 B^\prime},\mu\right)\right] \label{8}
\end{eqnarray}
where $\Delta_{B_1 B_2}$ is defined in Eq.~(\ref{5}). 

\subsection{Total correction}

The total correction to the baryon axial vector current matrix element from
Figs.~\ref{fig:axial}(a,b,c) is
\begin{eqnarray}
\left[\delta A^{ia}\right]_{B^\prime B} &=& 
\left[\delta A^{ia}\right]^{\rm vertex}_{B^\prime B}+{1\over 2}\left\{ 
\sum_{B_1}
 Z_{B^\prime B_1} \left[A^{ia}\right]_{B_1 B}+\sum_{B_2}
\left[A^{ia}\right]_{B^\prime B_2}Z_{B_2 B}\right\}\nonumber \\
&=& -\sum_{j,b,B_1,B_2} \left[A^{jb} \right]_{B^\prime B_2}
\left[A^{ia} \right]_{B_2 B_1}\left[A^{jb} \right]_{B_1 B}
{F\left(m_b,\Delta_{B_1 B},\mu\right)-
F\left(m_b,\Delta_{B_2 B^\prime},\mu\right)\over  \Delta_{B_1 B}-  \Delta_{B_2 B^\prime} }\nonumber \\
&&+{1\over2}\sum_{j,b,B_1,B_2} \left[A^{jb} \right]_{B^\prime B_2}
\left[A^{jb} \right]_{B^\prime B_2}\left[A^{ia} \right]_{B_2 B_1}
{\partial F(m_b,\Delta_{B_2 B_1},\mu) \over \partial \Delta_{B_2 B_1}}\nonumber \\
&&+{1\over2}\sum_{j,b,B_1,B_2} \left[A^{ia} \right]_{B^\prime B_2}
\left[A^{jb} \right]_{B_2 B_1}\left[A^{jb} \right]_{B_1 B}
{\partial F(m_b,\Delta_{B_1 B},\mu) \over \partial \Delta_{B_1 B}} .
\label{11}
\end{eqnarray}
(In addition to
$\delta A^{ia}$ of Eq.~(\ref{11}), there are also the contributions of Fig.~\ref{fig:axial}(d)
and the low-energy constants, which are considered in Sec.~VI.) 
Eq.~(\ref{11}) includes the full dependence on $\Delta/m$
of the one-loop correction. We want to
rewrite this expression so that the large-$N_c$ cancellations are manifest.

In the limit that the octet and decuplet baryons are degenerate, all the mass
differences $\Delta_{AB} \to 0$, and
\begin{equation}\label{10}
{1\over  \Delta_{B_1 B}-  \Delta_{B_2 B^\prime} }\left[ F\left(m_b,\Delta_{B_1 B},\mu\right)-
F\left(m_b,\Delta_{B_2 B^\prime},\mu\right)\right] \to F^{(1)}(m_b,0,\mu),
\end{equation}
where $F^{(n)}$ is defined by
\begin{equation}
F^{(n)}(m_b,\Delta,\mu)\equiv { \partial^n F(m_b,\Delta,\mu) \over \partial
\Delta^n}.
\end{equation}
In this limit, the correction to the axial current Eq.~(\ref{11}) reduces to
\begin{eqnarray}
\left[\delta A^{ia}\right]_{B^\prime B} &=& \sum_{j,b,B_1,B_2}F^{(1)}(m_b,0,\mu) 
\nonumber \\
&&\Biggl\{
-\left[A^{jb} \right]_{B^\prime B_2}
\left[A^{ia} \right]_{B_2 B_1}\left[A^{jb} \right]_{B_1 B} \nonumber\\
&&+ {1\over 2} \left[A^{jb} \right]_{B^\prime B_2}\left[A^{jb}\right]_{B_2 B_1}
 \left[A^{ia}\right]_{B_1 B} + {1\over 2} \left[A^{ia}\right]_{B^\prime B_2}
 \left[A^{jb} \right]_{B_2 B_1}\left[A^{jb}\right]_{B_1 B}\Biggr\}\label{12}
\end{eqnarray}
Let us adopt the more compact notation that $A^{ia}$ represents a matrix with
matrix elements  $\left[A^{ia}\right]_{B^\prime B}$, and summation over
intermediate baryon states is denoted by matrix multiplication. Then
Eq.~(\ref{12}) can be written as
\begin{eqnarray}
\delta A^{ia} &=& -\sum_{b,j}F^{(1)}(m_b,0,\mu) 
\left\{- A^{jb} A^{ia} A^{jb}
+ {1\over 2} A^{jb}A^{jb}
A^{ia} + {1 \over 2} A^{ia} A^{jb} A^{jb}\right\} \nonumber \\
&=&  -{1 \over 2}\sum_{b,j}F^{(1)}(m_b,0,\mu) 
\left[A^{jb},\left[A^{jb},A^{ia}\right]\right], \label{13}
\end{eqnarray}
which is the double-commutator form originally derived in 
Ref.~\cite{Dashen:1993as}.

The loop integral in the degeneracy limit $\Delta \rightarrow 0$ reduces to
\begin{equation}
F^{(1)}(m_b,0,\mu) = -{1\over 16 \pi^2 f^2}m_b^2 \left({11 \over 3} + \ln {m_b^2
\over \mu^2} \right).
\end{equation}
The $\ln m_b/\mu$ term is non-analytic in the quark mass, and is called a
``chiral log.''  The constant ($11/3$) piece is analytic in the quark masses,
and has the same form as higher dimension terms in the chiral Lagrangian. The
constant term is scheme-dependent, but the chiral logarithm is universal.

We discuss the structure of the large-$N_c$ cancellations for the baryon axial
vector currents in the next two sections.  First, in Sec.~IV, the cancellations
are studied in the degeneracy limit.  The generalization to non-degenerate baryons is given in
Sec.~V.

\section{Large-$N_c$ cancellations: $\Delta/m_\pi=0$}\label{sec:cancel}

The large-$N_c$ cancellations in the degeneracy limit for the one-loop correction to the 
baryon axial vector current follow from the double-commutator form of
Eq.~(\ref{13}). The pion decay constant $f \propto \sqrt{N_c}$, so the function
$F^{(1)}(m_b,0,\mu)$ is of order $1/N_c$. Each axial vector current matrix element is
of order $N_c$ (recall that $g_A$ is of order $N_c$), so the correction $\delta
A^{ia}$ is naively of absolute order $N_c^2$, i.e.\ of order $N_c$ relative to
the tree-level value $A^{ia}$. The large-$N_c$ consistency conditions derived
in Ref.~\cite{Dashen:1993as} imply that the double commutator
$\left[A^{jb},\left[A^{jb},A^{ia}\right]\right]$ is of order $N_c$ rather the
naive order $N_c^3$, \emph{provided one sums over all baryon states in a
complete multiplet of the large-$N_c$ $SU(6)$ spin-flavor symmetry, i.e.\ over both the
octet and decuplet, and uses axial coupling ratios given by the large-$N_c$ spin-flavor 
symmetry.}

Before discussing the cancellation in the double commutator,  we first review
some necessary large-$N_c$ formalism.  The baryon matrix element of the axial
vector current in QCD can be expanded in a $1/N_c$ expansion in terms of $SU(6)$
spin-flavor
operators~\cite{Dashen:1993as,Dashen:1994jt,Luty:1994fu,Carone:1994dz,Dashen:1995qi},
\footnote{For recent
reviews of the large-$N_c$ spin-flavor symmetry, see Refs.~\cite{jannrev}.}
\begin{equation}
G^{ia} = q^\dagger {\sigma^i \over 2} {\lambda^a \over 2} q,\qquad
T^a = q^\dagger {\lambda^a \over 2} q,\qquad
J^{i} = q^\dagger {\sigma^i \over 2} q,\qquad \label{14}
\end{equation}
where $q$ and $q^\dagger$ are $SU(6)$ operators that create and annihilate
states in the fundamental representation of $SU(6)$, and $\sigma^i$ and
$\lambda^a$ are the Pauli spin and Gell-Mann flavor matrices. The lowest mass baryon
multiplet transforms under $SU(6)$ as a completely symmetric tensor with $N_c$
indices. For $N_c=3$, this representation decomposes under spin and flavor into
a spin-1/2 octet and a spin-3/2 decuplet. The baryon axial vector current $A^{ia}$ in
the large-$N_c$ limit has the form~\cite{Dashen:1995qi}
\begin{eqnarray}
A^{ia} = a_1 G^{ia} + \sum_{n=2,3}^{N_c} b_n \frac{1}{N_c^{n-1}} {\cal 
D}_n^{ia} + \sum_{n=3,5}^{N_c} c_n \frac{1}{N_c^{n-1}} {\cal O}_n^{ia} \,
, \label{eq:axcurr}
\end{eqnarray}
where the coefficients are of order one.  The operators ${\cal D}_n^{ia}$ are diagonal
operators with nonzero matrix elements only between states with the same spin,
and the operators ${\cal O}_n^{ia}$ are purely off-diagonal operators with nonzero matrix
elements only between states of different spin. The explicit forms for these
operators can be found in Ref.~\cite{Dashen:1995qi}. At the physical value $N_c
= 3$, Eq.~(\ref{eq:axcurr}) reduces to
\begin{eqnarray}
A^{ia} & = & a_1 G^{ia} + b_2 \frac{1}{N_c} J^i T^a + b_3 \frac{1}{N_c^2}
{\cal D}_3^{ia} + c_3 \frac{1}{N_c^2} {\cal O}_3^{ia} \, ,\label{eq:axcurr3}
\end{eqnarray}
where
\begin{eqnarray}
{\cal D}_3^{ia} & = & \{ J^i, \{J^j, G^{ja} \} \} \, ,\nonumber \\
{\cal O}_3^{ia} & = & \{ J^2, G^{ia} \} - \frac12 \{ J^i, \{J^j, G^{ja} \}
\} \, .\label{opdef}
\end{eqnarray}

The four conventional $SU(3)$ baryon axial couplings $F$, $D$,
$\mathcal C$ and $\mathcal H$ for the baryon octet and decuplet  
can be written as linear combinations of the coefficients $a_1$,
$b_2$, $b_3$ and $c_3$ of the $1/N_c$ expansion,
\begin{eqnarray}
\begin{array}{l}
\displaystyle
D = \frac12 a_1 + \frac16 b_3 \, , \\ [3mm]
\displaystyle
F = \frac13 a_1 + \frac16 b_2 + \frac19 b_3 \, , \\ [3mm]
\displaystyle
{\cal C} = - a_1 - \frac12 c_3 \, , \\ [3mm]
\displaystyle
{\cal H} = - \frac32 a_1 - \frac32 b_2 - \frac52 b_3 \, .
\end{array} \label{eq:cid}
\end{eqnarray}
The leading order prediction of large-$N_c$ QCD is obtained by dropping the
$1/N_c$ suppressed terms in Eq.~(\ref{eq:axcurr}), i.e.\ the 3-body operators
${\cal D}_3^{ia}$ and ${\cal O}_3^{ia}$.  The $G^{ia}$ operator gives $F=2D/3$,
${\mathcal C}=-2D$ and ${\mathcal H}=-3D$, so that the coupling ratios, but not
necessarily their absolute normalization, are those predicted by $SU(6)$
symmetry.  The 2-body operator $J^i T^a$ corrects these relations. The
correction is of relative order $1/N_c^2$ for pions.

The baryon matrix elements of $J^i$ for the low-lying baryons in the $SU(6)$
representation are of order unity. The $N_c$ dependence of matrix elements of
$G^{ia}$ and $T^a$ is more subtle, and depends on the particular component $a$
chosen, as well as on the initial and final state baryon~\cite{Dashen:1995qi}.
For the purposes of this paper, we will use the naive estimate that matrix
elements of $G^{ia}$ and $T^a$ are both of order $N_c$, which is the largest
they can be. We focus upon baryons with spins of order unity.
The $N_c$ counting rules are summarized as:
\begin{equation}\label{ncount}
G^{ia} \sim N_c,\qquad
T^a \sim N_c, \qquad
J^i \sim 1.
\end{equation}
The $1/N_c$ expansion of a baryonic matrix element can be written as an
expansion in powers of $G^{ia}/N_c$, $T^a/N_c$ and $J^i/N_c$. The counting
rules Eq.~(\ref{ncount}) show that each factor $J$ leads to a $1/N_c$
suppression factor.

We can now understand the origin of the large-$N_c$ cancellations in
Eq.~(\ref{13}). At leading order in $N_c$, the axial current operator $A^{ia}$
can be replaced by $a_1 G^{ia}$, and has matrix elements of order $N_c$. The
commutator $\left[A^{ia},A^{jb}\right]= \left[a_1 G^{ia},a_1 G^{jb}\right]$ is
naively of order $N_c^2$, since each $G^{ia}$ is of order $N_c$. However, the
commutation relation
\begin{equation}\label{ggcomm}
\left[G^{ia},G^{jb}\right]={i\over 4} \delta^{ij} f^{abc} T^c + {i \over 6}
\delta^{ab} \epsilon^{ijk} J^k + {i\over 2} d^{abc} \epsilon^{ijk} G^{kc},
\end{equation}
shows that matrix elements of the commutator $\left[G^{ia},G^{jb}\right]$ are
at most of order $N_c$, since the right-hand side of Eq.~(\ref{ggcomm}) is at
most of order $N_c$. Thus there is a factor of $N_c$ cancellation between the
various terms in the commutator $\left[G^{ia},G^{jb}\right]$ from the summation
over intermediate baryon states. Similarly, one finds that there is a factor of
$N_c$ cancellation in the sum
over intermediate states for the commutators
\begin{equation}\label{tgcomm}
\left[T^a,G^{ib}\right]=i f^{abc} G^{ic},
\end{equation}
and
\begin{equation}\label{ttcomm}
\left[T^a,T^b\right]=i f^{abc} T^c,
\end{equation}
where the naive counting rule Eq.~(\ref{ncount}) has been used to estimate the 
order in $N_c$ of both sides of these equations. The basic reason for the
cancellation is that the maximum order in $N_c$ an $r$-body operator matrix
element can be is $N_c^r$, (an $r$-body operator is one with $r$ $q$'s and $r$
$q^\dagger$'s, i.e. can be written as a polynomial of order $r$ in $J^i$,
$G^{ia}$ and $T^a$), but the commutator of an $r$-body and $s$-body operator is
at most an $r+s-1$ body operator. Thus, every commutator potentially leads to a
cancellation by one factor of $N_c$. However, not every commutator gives a
factor of $N_c$ cancellation. The commutators
\begin{equation}\label{jjcomm}
\left[J^i,J^j\right]=i \epsilon^{ijk} J^k,
\end{equation}
and
\begin{equation}\label{jgcomm}
\left[J^i, G^{ja} \right]=i \epsilon^{ijk} G^{ka},
\end{equation}
have no cancellations, since both sides are of order one and order $N_c$,
respectively. The reason that there is no cancellation in Eqs.~(\ref{jjcomm})
and (\ref{jgcomm}) is that $J^i$ is a one-body operator whose matrix elements
are of order unity, rather than of order $N_c$.

Equations~(\ref{tgcomm})--(\ref{jgcomm}) lead to the conclusion that each
commutator produces a $N_c$ cancellation, unless a factor of $J^i$ is eliminated. 
The double-commutator in Eq.~(\ref{13}) has a
cancellation of $N_c^2$, because
$\left[G^{jb},\left[G^{jb},G^{ia}\right]\right]\sim J+G+T$, so that the
double-commutator is order $N_c$, rather than $N_c^3$. This was the
cancellation observed numerically in Ref.~\cite{Jenkins:1991jv}, and later proven in the $1/N_c$ expansion
in Refs.~\cite{Dashen:1993as,Dashen:1994jt}.

\section{Large-$N_c$ cancellations: $\Delta/m_\pi \not=0$}\label{sec:nonzero}

In this section, we analyze the large-$N_c$ cancellations in the renormalization
of the baryon axial vector current for finite $\Delta/m_\pi$.

Equation~(\ref{11}) can be expanded in a power series in $\Delta$. Expanding
the function $F(m,\Delta,\mu)$ in a power series, and collecting terms gives
\begin{eqnarray}
&&\delta A^{ia}=\sum_{j,b}\Biggl\{
-{1 \over 2} F^{(1)}(m_b,0,\mu) \left[ A^{jb}, \left[ A^{jb}, A^{ia} \right] \right]
+{1 \over 2} F^{(2)}(m_b,0,\mu) \left\{ A^{jb}, \left[ A^{ia}, \left[ {\cal M}, A^{jb} \right]
\right] \right\}\label{15} \\
&&+{1 \over 6} F^{(3)}(m_b,0,\mu) \left( -\left[ A^{jb}, 
\left[\left[{\cal M}, \left[ {\cal M}, A^{jb} \right]\right], A^{ia} \right] \right]
+ {1 \over 2} \left[ \left[ {\cal M}, A^{jb} \right], \left[\left[{\cal M}, A^{jb} \right],
 A^{ia} \right] \right]
\right)+\ldots \Biggr\}
\nonumber
\end{eqnarray}
where ${\cal M}$ is the baryon mass matrix. In deriving this result we have
converted explicit sums over intermediate baryons to implicit sums in the
matrix multiplications. One can use either the baryon mass matrix ${\cal M}$,
or the baryon mass-splitting matrix $\Delta {\cal M}$ in Eq.~(34),  since
${\cal M}$ differs from $\Delta {\cal M}$ by the average baryon mass times the
unit matrix, which commutes and drops out of Eq.~(\ref{15}).  To evaluate
Eq.~(\ref{15}) to all orders in $\Delta/m_\pi$ would be extremely difficult,
since one would have to sum an infinite series, with each term having a
coefficient which is a complicated commutator/anticommutator of $\cal M$'s and
$A^{ia}$'s.

We would like to evaluate graphs in heavy baryon chiral perturbation theory so that
the $1/N_c$ cancellations are manifest, and do not occur as numerical
cancellations at the end of the calculation. We will now show that the large
$N_c$ cancellations only occur in the first few terms of Eq.~(\ref{15}), so
that the remaining terms can be summed using conventional heavy baryon chiral
perturbation theory in the usual manner.

The expansion Eq.~(\ref{15}) has a different structure depending on whether one
has an even or odd number of insertions of the baryon mass operator ${\cal M}$.
Terms with $2r$ insertions of ${\cal M}$ have $2r+2$ commutators, whereas terms
with $2r+1$ insertions of ${\cal M}$ have $2r+2$ commutators and one
anticommutator.

The general form of the baryon mass operator in the $1/N_c$ expansion in the
$SU(3)$ limit is~\cite{Dashen:1993as,Dashen:1994jt,Luty:1994fu,Carone:1994dz,Dashen:1995qi}
\begin{equation}\label{mexp}
{\cal M} = N_c\left[ m_0+ m_1 {J^2\over N_c^2}+m_2 {{(J^2)^2}\over N_c^4}+\ldots\right].
\end{equation}
The importance of a given term in the $1/N_c$ expansion can be obtained by
counting powers of $J$. Each factor of $J/N_c$ leads to a $1/N_c$ suppression,
since $J$ is of order unity according to the counting rules Eq.~(\ref{ncount}).

We now have all the necessary ingredients to count the power in $1/N_c$ of a
general term in Eq.~(\ref{15}). The operators $A^{ia}$ and $\cal M$ are
one-body operators, with naive order $N_c$, so the ${\cal M}^r$ term in the
expansion in Eq.~(\ref{15}) is naively of order $N^{3+r-1}$, including the
factor of $1/N_c$ from the $1/f^2$ in the loop integral $F$, as shown in row
(C) of Table~\ref{tab:1}. 
\begin{table}
\caption{Table of the order in $N_c$ of the terms in the expansion of the
one-loop correction to the axial currents. See the text for an explanation of
the entries.
\label{tab:1}} 
$$
\arraycolsep 5pt
\begin{array}{ll|cccccccc}
\hline\hline
(A) & \hbox{number of ${\cal M}$'s} & 0 & 1 & 2 & 3 & 4 & 5 & 6 \\ 
(B) & \hbox{$m_q$ dependence} & m_q \ln m_q & m_q^{1/2} & \ln m_q & m_q^{-1/2} & 
m_q^{-1} & m_q^{-3/2} & m_q^{-2}\\
(C) & \hbox{naive $N_c$ power} & 2 & 3 & 4 & 5 & 6 & 7 & 8\\
(D) & \hbox{commutators} & 2 & 2 & 4 & 4 & 6 & 6 & 8\\
(E) & \hbox{net power} & 0 & 1 & 0 & 1 & 0 & 1 & 0\\
(F) & \hbox{number of $J$'s} & p & p+2 & p+4 & p+6 & p+8 & p+10 & p+12
\\[10pt]
(G) & \hbox{$J$'s left} & \cases{ 0 & $p=0,1$ \cr  p-2 & $p \ge 2$\cr} 
& p & p & p+2 & p+2 & p+4 & p+4\\[20pt]
\hline
& & & & & \\[-10pt]
(H) & \hbox{final $N_c$ power} & \cases{ 0 & $p=0,1$ \cr  2-p & $p \ge 2$\cr}
& 1-p^* & -p & -1-p & -2-p & -3-p & -4-p\\[10pt]
(I) & \hbox{usual $N_c$ power} & 2-p &
1-p & -p & -1-p & -2-p & -3-p & -4-p\\[10pt]
\hline\hline
\end{array}
$$
\hbox{${}^*$ Actually is 0 for $p=0$. See text.}
\end{table}%
The number of commutators in each term is listed in the next line in this
table. Every commutator (naively) leads to a decrease in the naive $N_c$ order by
unity, since the commutator of an $r$-body and an $s$-body operator is at most
an $r+s-1$ body operator. This leads to the $N_c$ power given in row (E).
Finally, we need to count the powers of $J$ in each term. Each factor of ${\cal
M}$ has at least two $J$'s, since the $m_0 N_c$ term in Eq.~(\ref{mexp}) drops
out of the expression Eq.~(\ref{15}). Each factor of $A^{ia}$ can have $\ge 0$
$J$'s, as is clear from Eqs.~(\ref{eq:axcurr})--(\ref{opdef}). The number of
$J$'s in the original expression is listed in row (F), where $p\ge0$ is the
number of extra $J$'s from $\cal M$ or $A^{ia}$, beyond the minimum values of
two and zero, respectively. Finally, note that each commutator can be used to
eliminate one power of $J$. Thus the net power of $J$ left is given by
subtracting the number of commutators from the number of $J$'s. The minimum
number of $J$'s is non-negative, and is listed in row (G). Thus, the final $N_c$
power (row (H)) is given by the net power in row (E) minus the minimum number
of $J$'s in row (G) since there is an additional $1/N_c$ factor for each $J$. 
One can compare this with the ``usual'' $N_c$ counting rule listed in row (I).
The usual counting rule is obtained by including a factor of $N_c$ for each
$A^{ia}$ (i.e. for each factor of $F$, $D$, $\cal C$ or $\cal H$), a factor of
$1/N_c$ for the $1/f^2$, and a factor of $1/N_c$ for each power of $\Delta$,
with $p\ge0$ representing $1/N_c$ suppressed terms.

One interesting point can be noted from Table~\ref{tab:1}. The dominant $1/N_c$
corrections from the baryon mass splittings are due to multiple insertions of
the  $J^2$ term in the baryon mass matrix. Two insertions of the $J^2$ term
(the $p=0$ term in the ${\cal M}^2$ column) is $N_c$ more important than one
insertion of the $J^4$ term (the $p=2$ term in the ${\cal M}^1$
column).

There is an extra cancellation in the term linear in ${\cal M}$ that is not
apparent in Table~\ref{tab:1}. We will discuss this new large-$N_c$
cancellation momentarily.  Including this effect, one sees from
Table~\ref{tab:1} that all terms in the expansion of Eq.~(\ref{11}) with two or
more powers of ${\cal M}$ have the same $N_c$ behavior as one finds with the
usual $N_c$ counting, i.e. these terms have no extra cancellations. One can
therefore treat all terms with two or more powers of ${\cal M}$ by conventional
heavy baryon chiral perturbation theory---compute all the graphs, with vertices
written in terms of $F$, $D$, $\cal C$ and $\cal H$. The only terms that have
to be treated specially are those with zero or one power of ${\cal M}$. To compute
graphs in the conventional way omitting the first two terms in Eq.~(\ref{15}) 
is trivial; one
simply rewrites the loop integral Eq.~(\ref{Fdef}) by explicitly extracting the
first three terms in an expansion in $\Delta$,
\begin{equation}
F(m_b,\Delta,\mu) = F(m_b,0,\mu)+F^{(1)}(m_b,0,\mu)\Delta +
{1\over 2} F^{(2)}(m_b,0,\mu)\Delta^2 + \tilde F(m_b,\Delta,\mu).
\end{equation}
and takes the standard expressions for the loop corrections written in terms of
$F$, $D$, $\cal C$ and $\cal H$, with $F \to \tilde F$. This procedure sums the
entire series in $\left(\Delta/m_\pi\right)^r$ starting with $r=3$. To this
result is added the first two terms in Eq.~(\ref{15}). One needs to extract
three terms from $F$ to obtain the first two terms in Eq.~(\ref{15}), since
since $F(m_b,0,\mu)$ cancels out of the correction. 

One can now analyze the first two terms in the expansion of Eq.~(\ref{11}),
which are given in Eq.~(\ref{15}). The first term is the double commutator term
discussed in the previous section. We see from Table~\ref{tab:1} that this term
is naively of order $N_c^2$, but actually is at most of order $N_c^0$. This is
consistent with the loop expansion being an expansion in $\hbar/N_c$, since the
one-loop correction is of order $1/N_c$ relative to the tree-level contribution
of order $N_c$. It is also apparent from Table~\ref{tab:1} that all terms of
order ${\cal M}^0$ with $p=0,1,2$ are equally important. Since there are no
powers of the baryon mass operator, the $p$ factors of $J$ must all arise from
$1/N_c$ corrections in the axial vertices $A^{ia}$. The expression for the
axial current relevant for $N_c=3$ is given in Eq.~(\ref{eq:axcurr3}), from
which it follows that terms with $p=0,1,2$ in the product $AAA$ are of the form
$GGG$,  $GG\, JT$, $G\,JT\, JT$, $G G{\cal D}_3$, and $G G{\cal O}_3$. All
these terms contribute {\bf at the same order} to the double-commutator,
whereas according to the usual counting one would have expected the $p=0$
product  $GGG$ to be one power of $N_c$ more important that the $p=1$ product
$GGJT$,  which in turn would be more important by one power of $N_c$ than the
$p=2$ products. This result has an important consequence: the one-loop
correction is very sensitive to the deviations of the axial coupling ratios
from their $SU(6)$ values. While the deviations are small corrections to the
couplings themselves, their importance gets enhanced in the one-loop
coefficient, because the leading term (proportional to $a_1^3$) is $1/N_c^2$
suppressed. Thus, for example, the $a_1$ term is the dominant contribution to
$F$, $D$, $\cal C$ and $\cal H$, and the $c_3$ term is a $1/N_c^2$ correction,
but the $a_1^3$ and $a_1^2 c_3$ terms are just as important in the one-loop
correction. Explicit forms for the one-loop correction in terms of $a_1$,
$b_2$, $b_3$ and $c_3$ will be given elsewhere.

The second term in Eq.~(\ref{15}) is
\begin{equation}
{1 \over 2} F^{(2)}(m_b,0,\mu) \left\{ A^{jb}, \left[ A^{ia},
\left[ {\cal M}, A^{jb} \right]
\right] \right\} \label{20}
\end{equation}
and is at most of order $N_c$ (using the value $(1-p)$ for $p=0$), 
the same order as the tree-level contribution to the
axial current. This result is surprising, because the $1/N_c$ expansion is a
semiclassical expansion in $\hbar/N_c$. One should be able to obtain the
leading in $1/N_c$ contributions from classical field theory. For example, it
was shown that the $N_c m_q^{3/2}$ one-loop correction to the baryon mass could
be obtained from the energy of the pion cloud coupled to a classical baryon
source~\cite{Manohar:1994iz}.  The term in
Eq.~(\ref{20}) involves the baryon mass splitting, which is a
quantum effect. The order $N_c$ contribution to Eq.~(\ref{20}) comes from using
$A^{ia}=a_1 G^{ia}$, and $\Delta{\cal M}=m_2 J^2/N_c$,
\begin{equation}
{{a_1^3 m_2}\over 2 N_c} F^{(2)}(m_b,0,\mu) \left\{ G^{jb}, \left[ G^{ia},
\left[ J^2 , G^{jb} \right]
\right] \right\}.\label{21}
\end{equation}
The operator factor $\left\{ G^{jb}, \left[ G^{ia},\left[ J^2 , G^{jb} \right] 
\right] \right\}$ is naively of order $N_c^3$, which implies that the correction 
Eq.~(\ref{21}) is an order $N_c$
correction to the axial currents, since $F^{(2)}(m_b,0,\mu)$ is of order
$1/N_c$. However, an explicit computation of the operator product using the
identities in Ref.~\cite{Dashen:1995qi} gives
\begin{eqnarray}
\left\{ G^{jb}, \left[ G^{ia},
\left[ J^2 , G^{jb} \right] \right] \right\}&=&-\left\{J^2, G^{ia}
\right\}+{1\over 2}\left(N_f+N_c\right)
 J^i T^a -{1\over 2} \left(N_f-2 \right)G^{ia},\label{prod}
\end{eqnarray}
which is only of order $N_c^2$, using the
$N_c$-counting rules in Eq.~(\ref{ncount}). The order $N_c^3$ part of
Eq.~(\ref{prod}) vanishes, which is a new cancellation in the one-loop
correction to the axial vector current. Consequently, Eq.~(\ref{20}) is of order
$N_c^0$ rather than order $N_c$, and is consistent with being a quantum
correction.

\section{Other contributions}\label{sec:other}

We have computed the chiral logarithmic correction to the axial vector current from
Figs.~\ref{fig:axial}(a,b,c). There is also the contribution from
Fig.~\ref{fig:axial}(d), which is 
\begin{equation} \delta A^{ia} = -{1 \over 2} \sum_b
\left[T^b ,\left[T^b, A^{ia} \right] \right] I(m_b),
\end{equation} 
where
\begin{eqnarray}
I(m_b) &=& {i \over f^2} \int {d^4k \over (2 \pi)^4}
 {1 \over k^2-m^2_b}
 = {m^2_b \over 16\pi^2 f^2}\left( \ln m^2_b/\mu^2 - 1 \right).
\end{eqnarray}
This contribution is of order $1/N_c$ relative to the tree-level contribution,
and does not involve any cancellations between Delta and nucleon states.

In addition to the loop corrections, one has the contribution from low-energy
constants multiplying higher dimension operators in the heavy baryon chiral
Lagrangian. These terms are analytic in the quark mass $m_q$. The analytic
contributions from the chiral Lagrangian can be of order $N_c$, i.e. the same order
in $N_c$ as
the tree-level contribution.

\section{Conclusions}

We have shown how to rewrite loop corrections in heavy baryon chiral perturbation
theory so as to include the full dependence on the Delta-nucleon mass
difference, while at the same time including the cancellations that follow from
the large-$N_c$ spin-flavor symmetry of baryons. The treatment in this paper
has included the decuplet-octet mass difference, but neglected 
the $SU(3)$ splittings of the octet and decuplet baryons.  It is possible to generalize
our analysis by including the $SU(3)$ mass splittings in the baryon mass
operator $\cal M$.

The one-loop correction to the baryon axial vector currents 
is very sensitive to deviations of
the axial couplings from their $SU(6)$ symmetry ratios, since the correction
that depends only on the $SU(6)$ coupling ratios (the $GGG$ term) is suppressed
by $1/N_c^2$, and the first subleading correction (the $GGJT$ term) is suppressed
by $1/N_c$.  Thus, the normally second subleading terms with two powers of $J$ in the
axial currents are as important as these two contributions.  We also have found
a new cancellation in the one-loop correction to the baryon axial vector current in the 
term linear in the baryon mass splittings.

\begin{figure}
\epsfxsize=6cm
\hfil\epsfbox{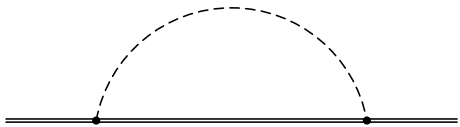}\hfill
\caption{One-loop correction to the baryon mass.
\label{fig:mass1}}
\end{figure}

\begin{figure}
\epsfxsize=12cm
\hfil\epsfbox{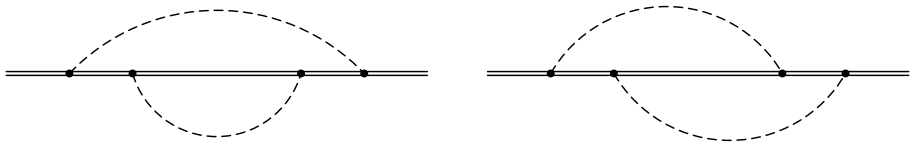}\hfill
\caption{Two-loop correction to the baryon mass.
\label{fig:mass2}}
\end{figure}

The large-$N_c$ cancellations play an important role in the one-loop
corrections to the axial vector current, and become more important at higher loops.
They also play an important role in the one-loop corrections to other baryon properties, 
such as the baryon
masses~\cite{Dashen:1993as,Dashen:1994jt,Jenkins:1995,Oh:1999yj}. 
At one loop, the $m_q^{3/2}$ correction to the baryon mass
from Fig.~\ref{fig:mass1}
is of order $N_c$, the same order as the tree-level baryon mass term, and there
is no cancellation between nucleon and Delta states. 
However, at two loops, the graphs
in Fig.~\ref{fig:mass2} produce $m_q^{5/2}$ corrections to the baryon mass 
that are formally of order $N_c^2$, but have cancellations which make the net correction
of order one.

\acknowledgements

This work was supported in part by the Department of Energy under Grant No.
DOE-FG03-97ER40546. R.F.M. was supported by CONACYT (Mexico) under the
UC-CONACYT agreement of cooperation and by CINVESTAV (Mexico). C.P.H.
acknowledges support from the Schweizerischer Nationalfonds and
Holderbank-Stiftung.  E.J. was supported  in part by the Alfred P. Sloan
Foundation and by the National Young Investigator program  through Grant No.
PHY-9457911 from the National Science Foundation.

\end{document}